\documentclass[pdflatex,sn-mathphys-num]{sn-jnl}

\usepackage{graphicx}%
\usepackage{multirow}%
\usepackage{amsmath,amssymb,amsfonts}%
\usepackage{amsthm}%
\usepackage{mathrsfs}%
\usepackage[title]{appendix}%
\usepackage{xcolor}%
\usepackage{textcomp}%
\usepackage{manyfoot}%
\usepackage{booktabs}%
\usepackage{algorithm}%
\usepackage{algorithmicx}%
\usepackage{algpseudocode}%
\usepackage{listings}%

\usepackage{siunitx} 
\usepackage{tikz}

\theoremstyle{thmstyleone}%

\theoremstyle{thmstyletwo}%

\theoremstyle{thmstylethree}%

\begin{document}

\title[Article Title]{Closed-loop control of a monolithically 3D nano-printed electromagnetic lens scanner with an integrated Hall sensor}

\author*[1]{\fnm{Florian} \sur{Lux}}\email{florian.lux@imtek.uni-freiburg.de}

\author[1]{\fnm{Elijah} \sur{Ditchendorf}}\email{elijah.ditchendorf@email.uni-freiburg.de}

\author[1]{\fnm{\c{C}a\u{g}lar} \sur{Ataman}}\email{caglar.ataman@imtek.uni-freiburg.de}

\affil*[1]{\orgdiv{Department}, \orgname{University of Freiburg}, \orgaddress{\street{Georges-Köhler-Allee 102}, \city{Freiburg}, \postcode{79110}, \country{Germany}}}

\abstract{

3D nano-printing through two-photon polymerization enables monolithic manufacturing of mechanical and freeform micro-optical elements with high inherent alignment accuracy. However, viscoelasticity and temperature-dependent stiffness of the photopolymer lead to hysteresis and drift, which significantly degrade open-loop position accuracy and long-term stability in quasi-static operation. Therefore, closed-loop control with integrated displacement sensing is essential for 3D nano-printed optical microsystems in practical precision positioning applications. Here, we present a closed-loop control system for a such a lens actuator that uses a commercial 3-axis Hall sensor for position tracking. A NdFeB micro-magnet encircling the integrated microlens provides both the actuation force and a position-dependent sensing signal. The Hall sensor, located between an anti-Helmholtz-like coil pair that drives the scanner bidirectionally, measures the combined field due to the micro-magnet and the coil pair. Calibration-based subtraction separates the coil field and sensor offset contributions to recover the magnet field and thus the axial magnet displacement. Closed-loop operation yields a mean absolute accuracy of \SI{0.86}{\micro\meter} and a precision of \SI{0.49}{\micro\meter} over a displacement range of \SI{150}{\micro\meter} while eliminating viscoelastic creep, suppressing hysteresis, and minimizing temperature-induced displacement drift under coil self-heating. This sensing approach requires no additional microfabrication steps and provides a practical path toward stable and repeatable positioning for monolithically 3D nano-printed optical microsystems.

}


\maketitle

\section{Introduction}\label{introduction}
Three-dimensional (3D) nano-printing through two-photon polymerization, also referred to as direct laser writing, is emerging as an enabling tool for the fabrication of micro- and millimeter-scale actuators, offering unprecedented optical and mechanical design freedom. By enabling the monolithic fabrication of mechanical, optical, and functional elements, this approach eliminates assembly-induced misalignment issues and allows rapid iteration of complex optical microsystems (MOEMS) \cite{Rothermel2024, Rothermel2025, Lux2025, Aybuke2024, Wende2025}. These actuators have demonstrated stroke ranges matching that of conventional microsystems in addition to the seamless integration of optical components, making them particularly attractive for miniaturized imaging systems \cite{Lux2025,Rothermel2024}, such as endomicroscopes \cite{Wende2025} and miniscopes \cite{Zong2022}, where size, weight, and alignment tolerances are of paramount importance. There is, however, an obvious caveat. 3D nano-printed actuators use photopolymers as structural material, whose intrinsic material qualities are far from those of monocrystalline silicon. The photopolymers commonly used for two-photon polymerization exhibit viscoelastic behavior, leading to creep and hysteresis \cite{Lux2025, Rothermel2024}. In addition, polymer aging and susceptibility to variations in environmental parameters, such as ambient temperature and humidity, further contribute to transient changes in mechanical properties \cite{Lux2025, Schmid_2009}. This is particularly critical for electromagnetic actuation, where Joule heating of the coils can cause a substantial temperature rise of the mechanical polymer structures, resulting in a reduction of the Young’s modulus \cite{ROHBECK2020108977} and, consequently, a drift in actuator response \cite{Lux2025}. All these factors render precise and repeatable open-loop operation impractical for many high-precision
applications. Thus, real-time monitoring and compensation of transient effects in the mechanical behavior of the 3D nano-printed optical microsystems through closed-loop control is a key step for their widespread adoption. 

The most important element of a closed-loop control system for microsystem devices is a reliable, compact, and integrated sensor for measuring the actuator displacement without compromising the size and alignment advantages of monolithic fabrication. Several sensing strategies compatible with micro- and nanoscale systems have been demonstrated in the context of 3D nano-printed actuators. Optical readout schemes, including interferometric or Fabry–Pérot–based sensing, have been successfully integrated into two-photon polymerized microactuators and microgrippers, enabling displacement or force estimation with high sensitivity \cite{Power2018, Thompson2018}. Similarly, fully 3D-printed scanning-probe systems have been realized that combine optical actuation with integrated optical displacement readout in a monolithic architecture \cite{Dietrich2020}. While these approaches offer excellent resolution, they typically require external optical interrogation, careful alignment, and fiber-based coupling, which complicates integration in compact actuators. Alternatively, strain-based sensing concepts, such as piezoresistive or strain-gauge transducers embedded into actuator structures or electrodes, have been used in MEMS and piezoelectric actuators to enable closed-loop displacement and force control \cite{Seethaler2021, Weng2025}. However, their implementation in polymer-based 3D nano-printed systems generally requires additional materials, lithographic post-processing, or electrical routing, thereby complicating the fabrication and assembly process. Capacitive sensing, which is the most common position sensing method for silicon microsystems \cite{ZHANGCapSensing}, either requires buried conductive elements within the printed structure or selective deposition of metal layers, both of which can be prohibitively difficult on a 3D nano-printed structure.

In this work, we discuss the implementation and performance of a closed-loop control system developed for a 3D nano-printed lens scanner recently demonstrated by our group \cite{Lux2025}. Bearing a large-aperture (\SI{1.4}{\milli\meter} diameter) aspherical lens, the device was designed and optimized for remote focusing in miniaturized imaging systems. It featured ortho-planar linear motion springs, a self-aligned sintered micro-magnet, and a monolithic lens, which was axially translated by a pair of micro-coils. The closed-loop system developed here uses a commercial 3-axis Hall sensor to track the magnetic field due to the micro-magnet \cite{MagneticPositionSensing}. The compact Hall sensor is accommodated within the actuator without increasing its footprint. Following a multi-step calibration process that quantifies the contributions of the micro-coil pair and the sensor offset, a single-axis field measurement can be related to the absolute magnet position through a simple, linear function. The position information is then used for a proportional-integral (PI) control loop, which achieves a mean absolute accuracy of \SI{0.86}{\micro\meter} and a precision of \SI{0.49}{\micro\meter} over a displacement range of \SI{150}{\micro\meter}, while effectively eliminating creep-induced displacement, suppressing hysteresis, and reducing temperature-induced displacement drift due to coil self-heating. 

\section{Results}\label{results}

\subsection{Concept and implementation}
\begin{figure}[htb]
\centering
\includegraphics[width=0.9\textwidth]{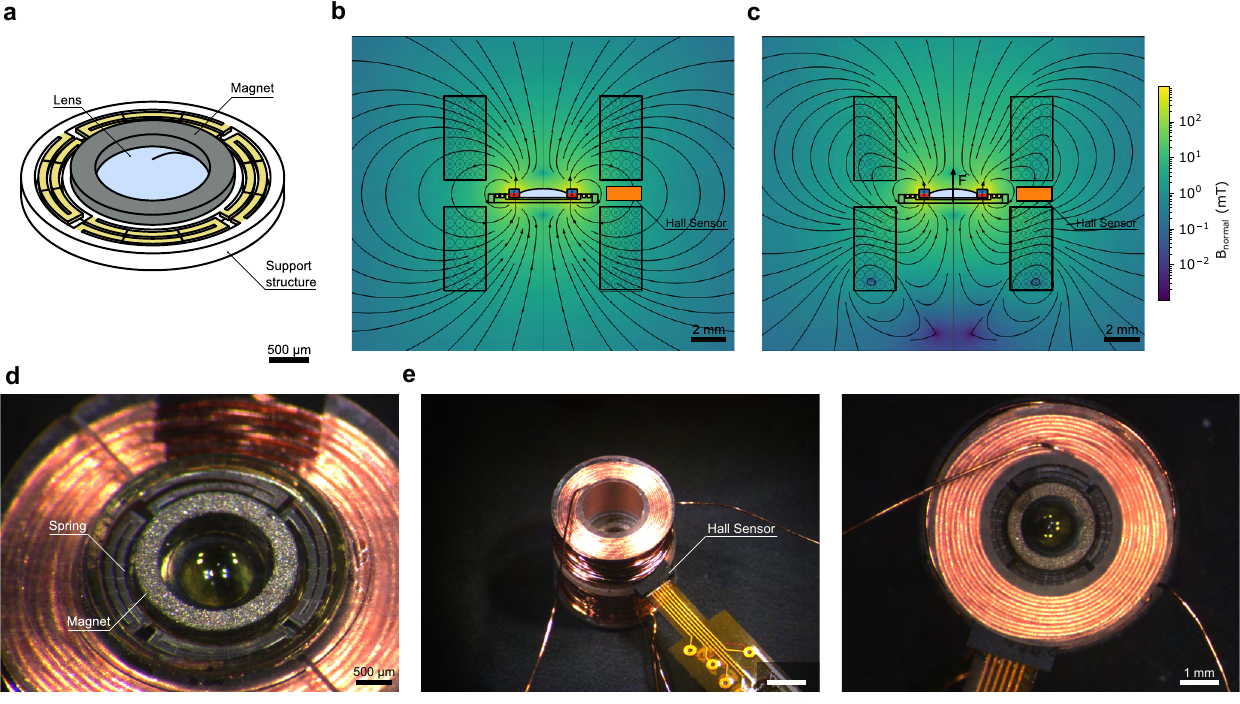}
\caption{\textbf{Working principle and photographs of the microlens scanner.} \textbf{a} Four ortho-planar linear-motion springs connect the shuttle, consisting of a magnet and an aspherical microlens, to the outer support structure. \textbf{b} To measure the displacement of the shuttle, the magnetic field created by the moving magnet is used. This field is measured using a Hall sensor sandwiched between the two coils. \textbf{c} Two coils in an anti-Helmholtz-like configuration create the magnetic field gradient required to displace the shuttle. The magnetic field of the micro-magnet becomes distorted (\(I_\textit{coil}\)\,=\,\SI{75}{\milli\ampere}). \textbf{d} Close-up photograph of the shuttle and the springs. \textbf{e} Fully assembled device, showing the Hall sensor mounted to a flex-PCB.}\label{fig:working_principle_pictures}
\end{figure}

The proposed mechanism for closed-loop control is demonstrated on recently-developed a 3D nano-printed lens scanner \cite{Lux2025} intended for active miniaturized optical instruments such as focus tuning in miniscopes \cite{Zong2021, Zong2022}. The design of the actuator is shown in Fig.\,\ref{fig:working_principle_pictures}\,(a). Inside the support structure are four ortho-planar linear-motion springs that connect to a movable shuttle. The shuttle carries a monolithically integrated aspherical microlens and a ring magnet with axial magnetization. Figure\,\ref{fig:working_principle_pictures}\,(d) shows a close-up photograph of the actuator. To displace the lens along the optical axis, the actuator is placed between two electromagnetic coils with opposite current directions, as shown in Fig.\,\ref{fig:working_principle_pictures}\,(b-c). The configuration of these two coils is comparable to that of an anti-Helmholtz coil. When driven with a current, this coil pair creates a large magnetic field gradient at the mid-plane between the coils, while minimizing the magnetic field itself. Since the micro-magnet is axially magnetized, it experiences a large force parallel to the optical axis and a minimal torque that can lead to tilt \cite{Lux2025, Aybuke2024}. A 3-axis 16-bit Hall sensor (MLX90394, Melexis, Belgium) with a high dynamic range (\SI{0.15}{\micro\tesla\per LSB} over a range of \textpm\,\SI{5}{\milli\tesla}) assembled onto a flexible printed circuit board (flex-PCB) is sandwiched between the two coils, as close to the magnet as possible. With total dimensions of 2\,x\,1.5\,x\,0.35\,mm\textsuperscript{3}, the sensor is compact enough to fit in between the coils, without increasing the device footprint. Photographs of the complete actuator with the integrated Hall sensor are depicted in Fig.\,\ref{fig:working_principle_pictures}\,(e). Figure\,\ref{fig:working_principle_pictures}\,(b) plots the normal magnetic field of the micro-magnet without any coil current while Fig.\,\ref{fig:working_principle_pictures}\,(c) plots the superposition of the magnetic field of the micro-magnet and the field of the coil for a current of \SI{75}{\milli\ampere}. The distortion of the magnetic field by the coil reveals that an isolation of the magnetic field of the micro-magnet from the field of the coil is essential for position sensing.

The magnetic field measured by the Hall sensor comprises multiple individual contributions: the offset \(\vec{B}_\textit{offset}\) of the Hall sensor, which includes static environmental (external) fields, the magnetic field \(\vec{B}_\textit{coil}\) of the coil, which is a function of coil current \(I_\textit{coil}\), and the magnetic field \(\vec{B}_\textit{magnet}\) of the magnet, which is a function of magnet position \(z_\textit{magnet}\) and thus shuttle/lens position. Assuming purely axial shuttle motion (i.e., negligible lateral displacement or tilt) and neglecting external time-dependent magnetic fields, the magnetic field recorded by the Hall sensor is given by

\begin{equation} \label{eq1}
\begin{split}
\vec{B} & = \vec{B}_\textit{offset}+\vec{B}_\textit{coil}+\vec{B}_\textit{magnet} \\
 & = \begin{bmatrix}
{B}_\textit{offset,x}\\
{B}_\textit{offset,y}\\
{B}_\textit{offset,z}\\
\end{bmatrix}+I_\textit{coil}\begin{bmatrix}
\frac{d B_x}{d I_\textit{coil}}\\
\frac{d B_y}{d I_\textit{coil}}\\
\frac{d B_z}{d I_\textit{coil}}\\
\end{bmatrix}+\begin{bmatrix} 
B_\textit{magnet,x}(z_\textit{magnet})\\
B_\textit{magnet,y}(z_\textit{magnet})\\
B_\textit{magnet,z}(z_\textit{magnet})\\
\end{bmatrix}.
\end{split}
\end{equation}

The magnet position \(z_\textit{magnet}\) can be calculated by subtracting the Hall sensor offset and the coil-generated field from the recorded field data. For this simple calculation to work, however, \(I_\textit{coil}\) should be provided by a low-noise driver, and the values of \(\vec{B}_\textit{offset}\), \(\frac{d\vec{B}}{dI_\textit{coil}}\) and \(\vec{B}(z_\textit{magnet})\) should be determined in a calibration routine. The next section describes the details of the calibration procedure. The real-time monitoring \(z_\textit{magnet}\) through this procedure enables the implementation of a proportional-integral (PI) control loop to precisely compensate for any deviations from linear mechanical behavior, e.g., thermal alteration of the Young's modulus and thus spring stiffness and viscoelastic creep. A PI controller was chosen over a PID controller because the sampling rate of the utilized Hall sensor (\SI{1.4}{\kilo\hertz}) is only about four times larger than the damped natural frequency of the device (\SI{362}{\hertz}). Including a derivative term at the given sampling frequency would cause instability in the control loop. The implementation of the control-loop is described in the Materials and Methods section.

\subsection*{Sensor calibration}\label{magnetic_field_characterization}
The coefficients required to distinguish the individual contributions to the measured magnetic field and to convert the corrected Hall sensor signal into an actuator displacement are obtained through a calibration procedure. As introduced in Eq.\,\ref{eq1}, the recorded magnetic field \(\vec{B}\) is modeled as the sum of the sensor offset \(\vec{B}_\textit{offset}\), the coil-induced field \(\vec{B}_\textit{coil}\), which scales linearly with coil current \(I_\textit{coil}\), and the field contribution of the magnet \(\vec{B}_\textit{magnet}\), which changes with its position \(z_\textit{magnet}\), and therefore shuttle/lens displacement. The calibration procedure therefore consists of three steps: determination of the sensor offset \(\vec{B}_\textit{offset}\), determination of the coil induced field as a function of current \(\frac{d\vec{B}}{dI_\textit{coil}}\), and determination of the field-to-displacement sensitivity \(\frac{d\vec{B}}{dz_\textit{magnet}}\) using a laser Doppler vibrometer (LDV) as a reference measurement.

\begin{figure}[htb]
	\centering
	\includegraphics[width=0.9\textwidth]{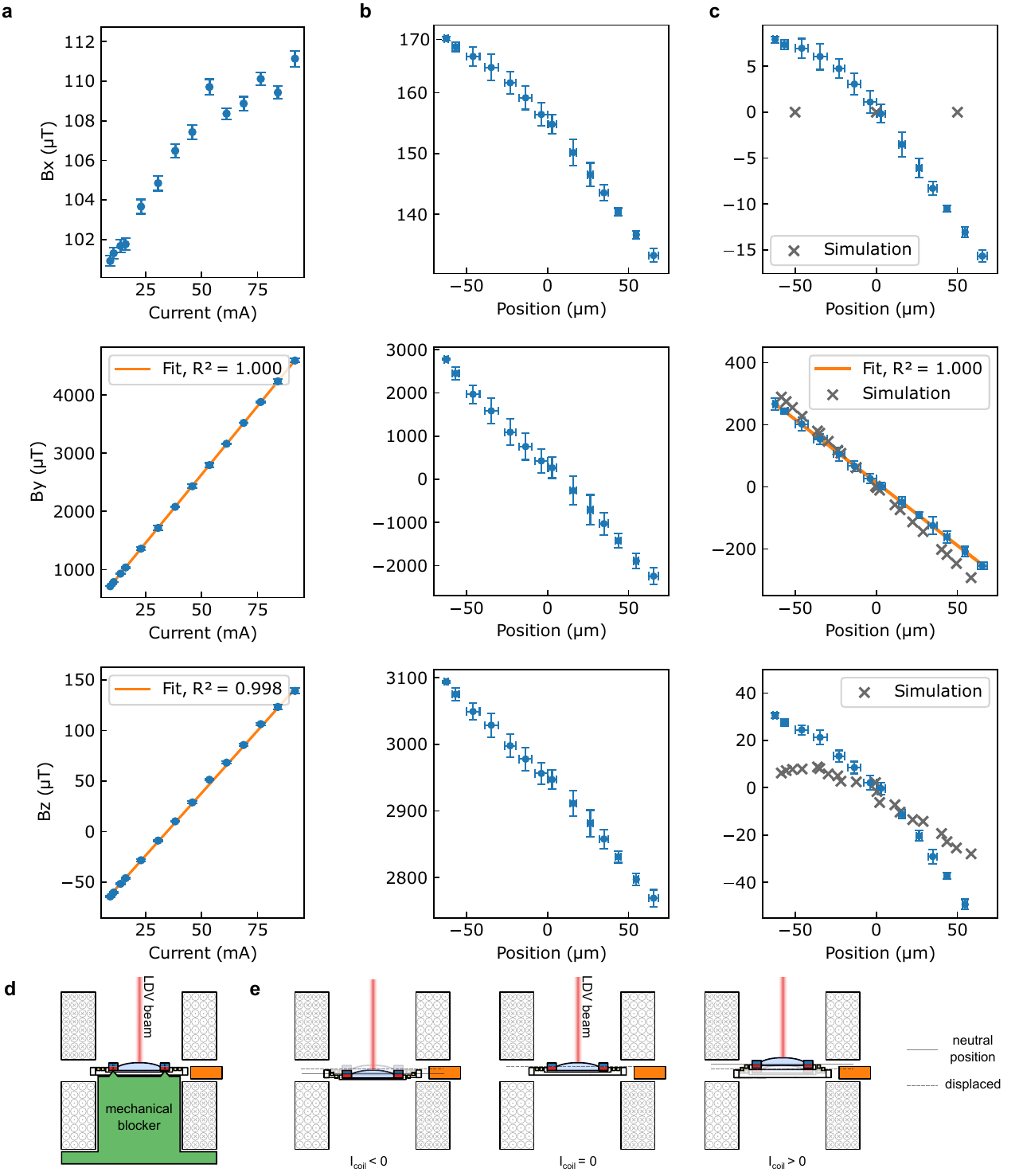}
	\caption{\textbf{Sensor calibration.} \textbf{a} Magnetic field as a function of coil current while the shuttle is fixed (n=100). \textbf{b} Magnetic field as a function of displacement without correcting for the magnetic field of the coil (n\textgreater 100). \textbf{c} Magnetic field as a function of displacement with correcting for the magnetic field of the coil (n\textgreater 100). \(B_y\) is used for sensing, as it shows the highest sensitivity and linearity. \textbf{d} The shuttle is mechanically blocked to isolate the coil contribution to the magnetic field from the contribution of the displaced magnet while calibrating the contribution of the coil to the magnetic field. \textbf{e} Characterization of the magnetic field contribution of the magnet as a function of displacement. Drawings not to scale.}\label{fig:coil_position-calibration}
\end{figure}

First, the sensor offset \(\vec{B}_\textit{offset}\) was measured at zero coil current (\(I_\textit{coil}=0\)). This offset captures the inherent sensor bias as well as any static field present. The resulting vector \(\vec{B}_\textit{offset}\) was stored and subsequently subtracted from all field measurements during closed-loop operation. Second, the magnetic field contribution of the coils was calibrated independently of the motion of the magnet. To suppress magnet displacement while still allowing a non-zero coil-current, a mechanical blocker was inserted into the coil assembly from the bottom until the shuttle motion was physically blocked, as shown in Fig.\,\ref{fig:coil_position-calibration}\,(d). The blocked state was verified using a LDV, ensuring that the actuator did not continue to move when the current was further increased. Due to the expected linear and antisymmetric relationship between coil current and magnetic field, measurements were limited to positive currents only. With the actuator immobilized, the coil current \(I_\textit{coil}\) was swept while recording the corresponding Hall sensor field vector \(\vec{B}(I_\textit{coil})\). Linear regression of \(\vec{B}(I_\textit{coil})\) data shown in Fig.\,\ref{fig:coil_position-calibration}(a) provided the coil sensitivity \(\frac{d\vec{B}}{dI_\textit{coil}}\) which quantifies the current-dependent field contribution of the coils at the sensor location. The recorded magnetic field scales linearly with the applied current. The slopes of \(B_y\) and \(B_z\) are \SI{47.1}{\micro\tesla\per\milli\ampere} and \SI{2.47}{\micro\tesla\per\milli\ampere} which closely match the simulated slopes of \SI{41.6}{\micro\tesla\per\milli\ampere}, and \SI{0.125}{\micro\tesla\per\milli\ampere}. \(B_x\) is altered by only \SI{10}{\micro\tesla}, matching the simulations, which predict that \(B_x\) is not affected by the coil pair.
Third, the relationship between the magnetic field and the magnet displacement was measured. After removing the mechanical blocker, the coil current \(I_\textit{coil}\) was swept to displace the magnet, as shown in Fig.\,\ref{fig:coil_position-calibration}\,(e), while simultaneously recording the magnetic field vector \(\vec{B}\) from the Hall sensor and the actuator position \(z_\textit{magnet}\) using the LDV as a reference measurement. To isolate the magnet-induced field variation from the total recorded field, the raw Hall sensor data were corrected by subtracting the offset and the calibrated coil contribution, i.e.,
\(\vec{B}_\textit{corr.}=\vec{B}-\vec{B}_\textit{offset}-I_\textit{coil}\frac{d\vec{B}}{dI_\textit{coil}}\).
The remaining field variation \(\vec{B}_\textit{corr.}\) is dominated by the permanent magnet and therefore correlates with the measured displacement \(z_\textit{magnet}\). 
Figure\,\ref{fig:coil_position-calibration} (b) shows the raw data, dominated by the contribution of the coil to the magnetic field, while the data in Fig.\,\ref{fig:coil_position-calibration} (c) are with the magnetic field contribution of the coil subtracted from the data. Due to its high sensitivity and linearity, \(B_y\) is used for sensing. The other fields are not used due to their non-linearity and because their sensitivity is about an order of magnitude lower. A linear regression was then used to determine \(\frac{d{B_y}}{dz_\textit{magnet}}\). The slope \(\frac{d{B_y}}{dz_\textit{magnet}}\) is \SI{-4.05}{\micro\tesla\per\micro\meter}, which is close to the simulated slope of \SI{-4.99}{\micro\tesla\per\micro\meter}. Deviations between measured and simulated values for \(\frac{d\vec{B}}{dI_\textit{coil}}\) and \(\frac{d\vec{B}} {dz_\textit{magnet}}\) are associated with a tilt of the Hall sensor due to alignment tolerances and inhomogeneities in the manual coil winding.

The magnetic characterization shows that the coil contribution is approximately an order of magnitude larger than the magnet contribution. This renders accurate calculation of the magnet position rather challenging, as any residual error in the subtraction of the coil field manifests itself as a position error. Because this subtraction relies on calibrated \(\frac{d\vec{B}}{dI_\textit{coil}}\), accurate current control is essential. The custom current source/sink used in this work has a maximum error of \SI{0.86}{\micro\ampere} over the calibrated range, which corresponds to a theoretical position readout error of only \textpm\SI{10}{\nano\meter} when mapped via \(\frac{dB_y}{dI_\textit{coil}}=\)\SI{47.1}{\micro\tesla\per\milli\ampere} and \(\frac{dB_y}{dz_\textit{magnet}}=\)\SI{-4.05}{\micro\tesla\per\micro\meter}.

\subsection*{Accuracy, precision and measurement range}\label{accuracy_precision}

\begin{figure}[htb]
\centering
\includegraphics[width=0.9\textwidth]{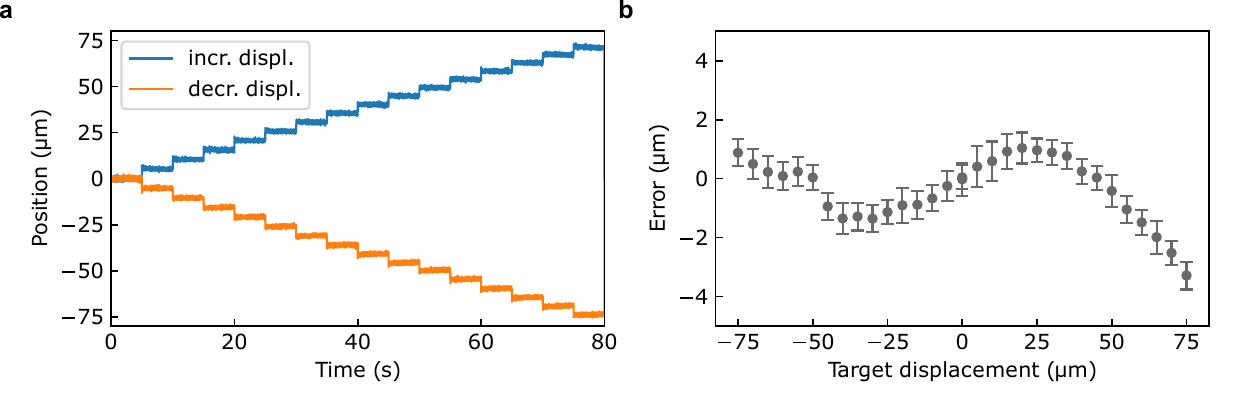}
\caption{\textbf{Accuracy and precision characterization data.} \textbf{a} The actuator was displaced in steps of \SI{5}{\micro\meter} in positive and negative direction. \textbf{b} For each level, the accuracy and precision (errorbars) were calculated (n=2500). The accuracy shows a systematic trend with a mean absolute accuracy of \SI{0.86}{\micro\meter}. The mean precision (std) is: \SI{0.49}{\micro\meter} without showing a systematic trend.}\label{fig:accuracy_precision}
\end{figure}

To measure the accuracy and the precision of the microlens scanner under closed-loop control, the target displacement was increased/decreased in steps of \SI{5}{\micro\meter} until a displacement of \textpm\,\SI{75}{\micro\meter} was reached, as shown in Fig.\,\ref{fig:accuracy_precision}\,(a). At each interval, 2500 position datapoints were collected using a LDV as a reference measurement. The mismatch between target displacement and displacement measured using the LDV as a function of displacement is shown in Fig.\,\ref{fig:accuracy_precision}\,(b). The error shows a systematic trend. The mean absolute accuracy of the closed-loop system is \SI{0.86}{\micro\meter}. The mean precision is \SI{0.49}{\micro\meter}.

Because the position estimate is obtained by subtracting the coil contribution and applying a linear field-to-position mapping, the most plausible origins of the observed error are residual calibration/model errors. Residuals in the coil calibration of up to \SI{5}{\micro\tesla} correspond to a position error of \SI{1.2}{\micro\meter}, and residuals in the position calibration of up to \SI{2}{\micro\tesla} correspond to a position error of \SI{0.5}{\micro\meter}. When applying a linear model to the simulated magnetic field, the maximum residual is \SI{1}{\micro\tesla}, corresponding to a displacement error of \SI{0.25}{\micro\meter}, justifying the use of a linear model. An estimate of the noise-limited displacement precision was obtained by assuming RMS white Hall-sensor noise of \(\sigma_b\) of \SI{3.5}{\micro\tesla} over the Nyquist band (0–\(f_s/2\)
, \(f_s\)=\SI{1.4}{\kilo\hertz}) and using the equivalent noise bandwidth of the applied 1\textsuperscript{st} order low pass filter in the control loop (\(\pi/2\)) with \(f_c\) = \SI{200}{\hertz}. Mapping the resulting field noise through the calibrated sensitivity \(\frac{dB_y} {dz_\textit{magnet}}\)\,=\,\SI{-4.05}{\micro\tesla\per\micro\meter} yields \(\sigma_z\)\,=\,\SI{0.58}{\micro\meter}, consistent with the measured precision.

The measurement range is ultimately limited by the range of the Hall sensor in Low Noise configuration (\textpm\,\SI{5}{\milli\tesla}). Since the magnetic field at the sensor location is dominated by the contribution of the coil, the Hall sensor becomes saturated at a current of \SI{106}{\milli\ampere}, assuming no external fields or offsets. This current corresponds to an estimated displacement of \textpm\,\SI{90}{\micro\meter}. Using the Hall sensor in the High Range configuration can overcome this limitation at the cost of a decrease in precision.

\subsection*{Step response}\label{step_response}

\begin{figure}[htb]
\centering
\includegraphics[width=0.9\textwidth]{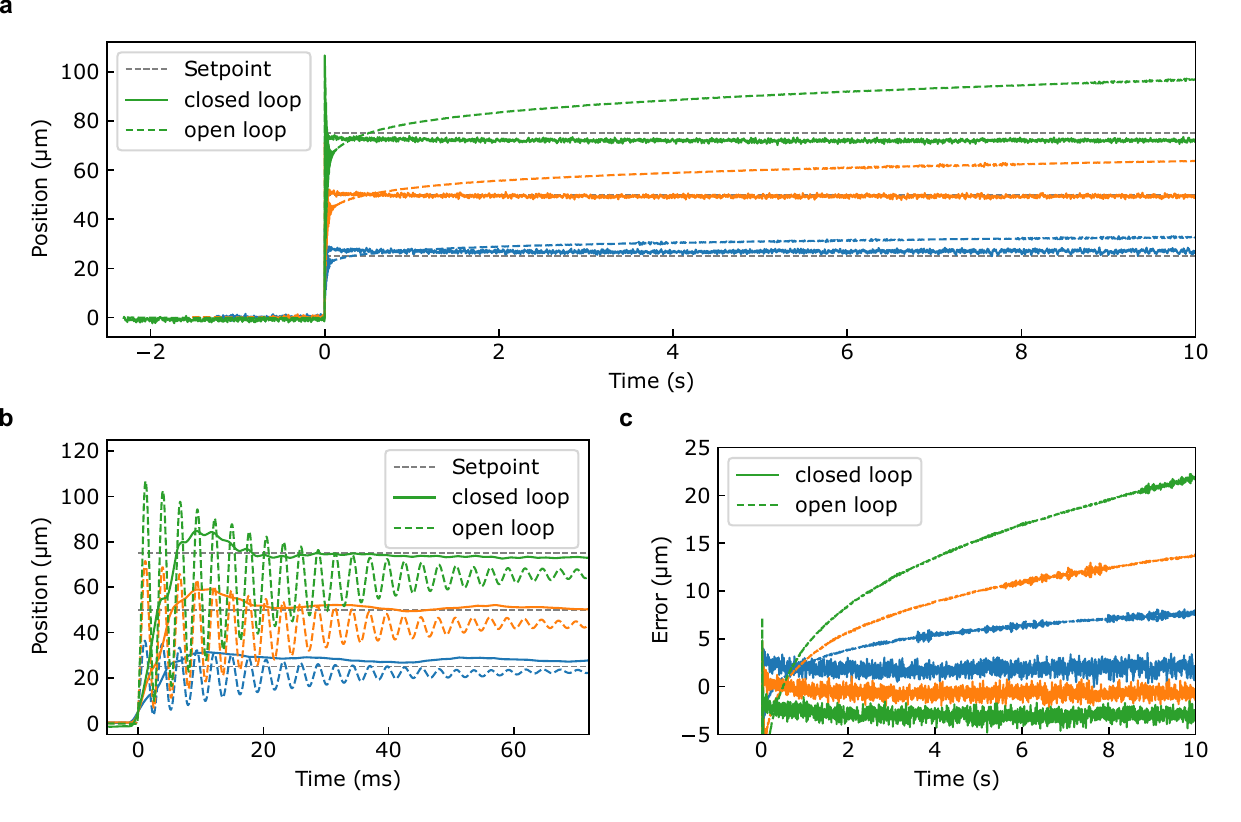}
\caption{\textbf{Open and closed-loop step response of the lens scanner.} \textbf{a} In open-loop operation, viscoelastic creep is present. In closed-loop operation, this creep is eliminated by the control-loop. \textbf{b} Following the step input, an overshoot of \SI{44.3}{\percent}\,\textpm\,\SI{2.8}{\percent} and ringing at the damped natural frequency can be observed for open-loop operation. In closed-loop operation, no ringing can be observed. The overshoot is reduced to \SI{18.8}{\percent}\,\textpm\,\SI{0.3}{\percent}. \textbf{c} Mismatch between setpoint and displacement, showing the viscoelastic drift in open-loop operation.} \label{fig:step_response}
\end{figure}

To characterize the system dynamics and investigate the extent to which the control system can account for the viscoelastic response of the polymer to a step input, the actuator was excited using step inputs at three different amplitudes (\SI{25}{\micro\meter}, \SI{50}{\micro\meter}, \SI{75}{\micro\meter}). Figure\,\ref{fig:step_response}\,(a) shows the measured displacement (LDV) over a range of \SI{10}{\second}. Due to the viscoelastic properties of the actuator, which have been characterized in detail in previous works from our group and others \cite{Lux2025,Rothermel2024}, a significant drift can be observed for the open-loop case. For a step input aiming at a displacement of \SI{75}{\micro\meter}, the measured drift after \SI{10}{\second} was \SI{22}{\micro\meter} (\SI{29}{\percent}). For the closed-loop, no significant drift can be observed. As shown in Fig.\,\ref{fig:step_response}\,(c) there is a difference between the target displacement and the measured displacement. This difference matches the displacement-dependent accuracy measurement depicted in Fig.\,\ref{fig:accuracy_precision}\,(b).

The settling time \(T_s\) (with \textpm\,\SI{5}{\percent} error band) in open-loop operation was \SI{69.7}{\milli\second}\,\textpm\,\SI{5.6}{\milli\second}. In closed-loop control, the settling time is reduced to \SI{17.6}{\milli\second}\,\textpm\,\SI{0.4}{\milli\second}. The overshoot was reduced from \SI{44.3}{\percent}\,\textpm\,\SI{2.8}{\percent} to \SI{18.8}{\percent}\,\textpm\,\SI{0.3}{\percent}. Furthermore, the ringing at the damped natural frequency related to the system's inherent natural frequency and damping ratio that can be observed in open-loop operation is reduced in closed-loop operation, as shown in Fig.\,\ref{fig:step_response}\,(b).

\subsection*{Hysteresis}\label{hysteresis}

\begin{figure}[htb]
\centering
\includegraphics[width=0.9\textwidth]{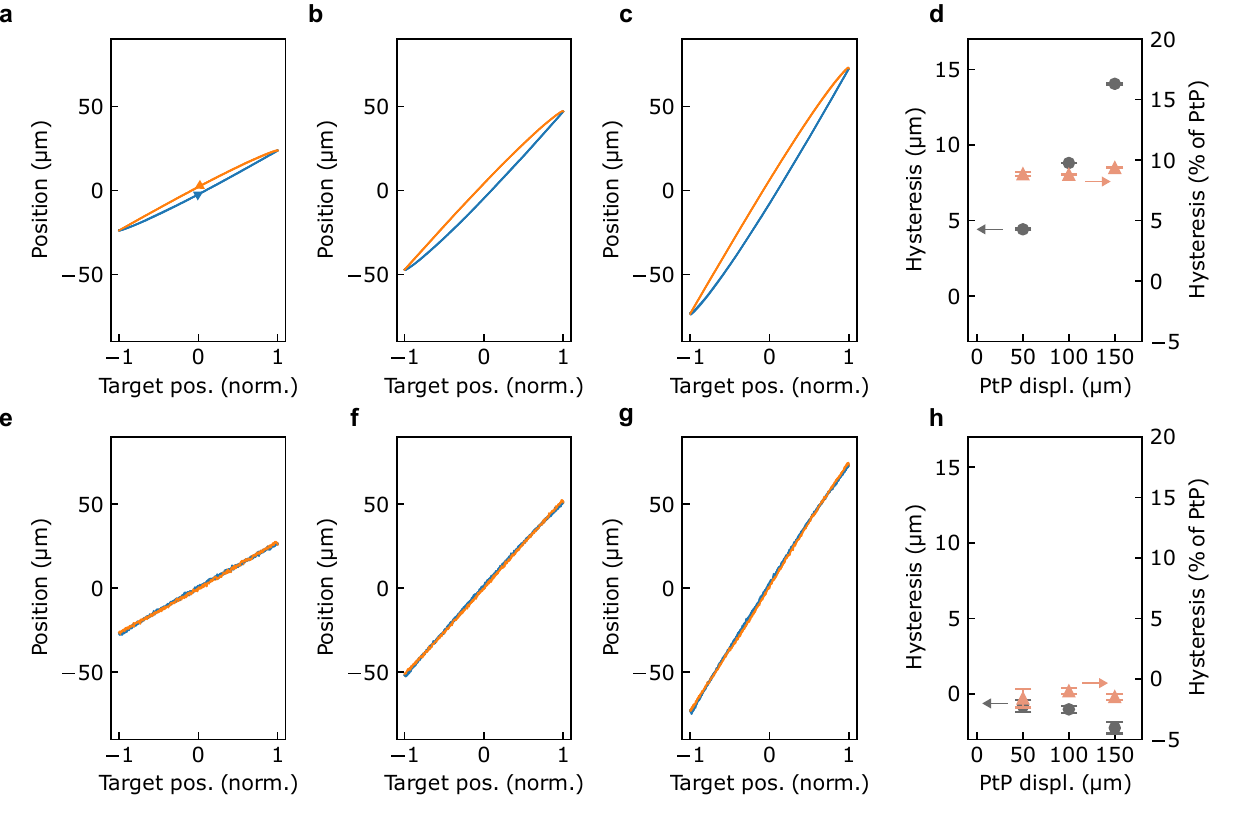}
\caption{\textbf{Hysteresis characterization of the actuator.} \textbf{a-c} Displacement as function of normalized target displacement for setpoints of \textpm\SI{25}{\micro\meter}, \textpm\SI{50}{\micro\meter}, and \textpm\SI{75}{\micro\meter}, showing hysteretic behavior due to combined elastic and viscoelastic contributions at an actuation frequency of \SI{0.25}{\milli\hertz}. The blue lines correspond to increasing displacement, orange lines to decreasing displacement \textbf{d} Hysteresis in open-loop operation as a function of peak-to-peak displacement (n=5). \textbf{e-g} Corresponding closed-loop measurements for the same setpoints, where forward and backward trajectories largely overlap, indicating effective hysteresis compensation. \textbf{h} Hysteresis in closed-loop operation as function of peak-to-peak displacement (n=5).}\label{fig:hysteresis}
\end{figure}

When polymer-based actuators are driven with a symmetric signal centered around zero at a frequency with a period shorter than the time constants of the viscoelastic creep, the displacement is dominated by elastic effects. The result is reduced hysteresis compared to actuation at lower frequencies with periods comparable to the time constants of viscoelastic creep. In this low frequency scenario, both the elastic and viscoelastic effects contribute to displacement, leading to a hysteretic behavior \cite{Sheng2018}. The time constants of the viscoelastic creep for the given device structure/material are \SI{3.08}{\second}\,\textpm\,\SI{0.22}{\second} and \SI{51.42}{\second}\,\textpm\,\SI{0.60}{\second} \cite{Lux2025} for a general Kelvin-Voigt viscoelasticity model of second order \cite{Serra2019}. To investigate the ability of the control system to compensate for the described effect, we drove the device with a triangular signal at \SI{0.25}{\milli\hertz} with varying setpoints (\textpm\SI{25}{\micro\meter}, \textpm\SI{50}{\micro\meter}, \textpm\SI{75}{\micro\meter}). 
Figures\,\ref{fig:hysteresis}\,(a-c) show the displacement as a function of normalized target displacement in open-loop operation. A hysteretic behavior is evident. The measured hysteresis is a function of target displacement, as shown in Fig.\,\ref{fig:hysteresis}\,(d). The mean relative hysteresis in open-loop operation is \SI{9.0}{\percent}\,\textpm\SI{0.3}{\percent}. In closed-loop operation, the hysteresis is reduced, as shown in Fig.\,\ref{fig:hysteresis}\,(e-g). The mean of the hysteresis, illustrated in Fig.\,\ref{fig:hysteresis}\,(h), is \SI{-1.4}{\percent}\,\textpm\,\SI{0.3}{\percent}. The negative sign is associated with the integral term of the PI controller, as at low frequencies, the controller dynamics are dominated by the integral action of the PI controller, which exhibits a \SI{-90}{\degree} phase contribution.

\subsection*{Temperature effects}\label{temperature_effects}
Even though the lens scanner relies on moving-magnet actuation, and thus, there is no direct heat dissipation on the moving structures, Joule heating due to the coils can still increase the device temperature. An increase in temperature leads to a heat-induced reduction of the Young's modulus, which in turn leads to an increase in mechanical response, i.e. \SI{}{\micro\meter\per\milli\ampere}. This effect manifests itself as mechanical drift during continuous operation. The temperature dependence of the Young's modulus for IP-Dip (Nanoscribe GmbH, Germany), a photoresin closely related to IP-S used in this work, was demonstrated by Rohbeck et al \cite{ROHBECK2020108977}, where a reduction in Young's modulus of more than \SI{5}{\percent} was observed for a temperature increase from \SI{20}{\celsius} to \SI{25}{\celsius}. 

\begin{figure}[htb]
\centering
\includegraphics[width=0.9\textwidth]{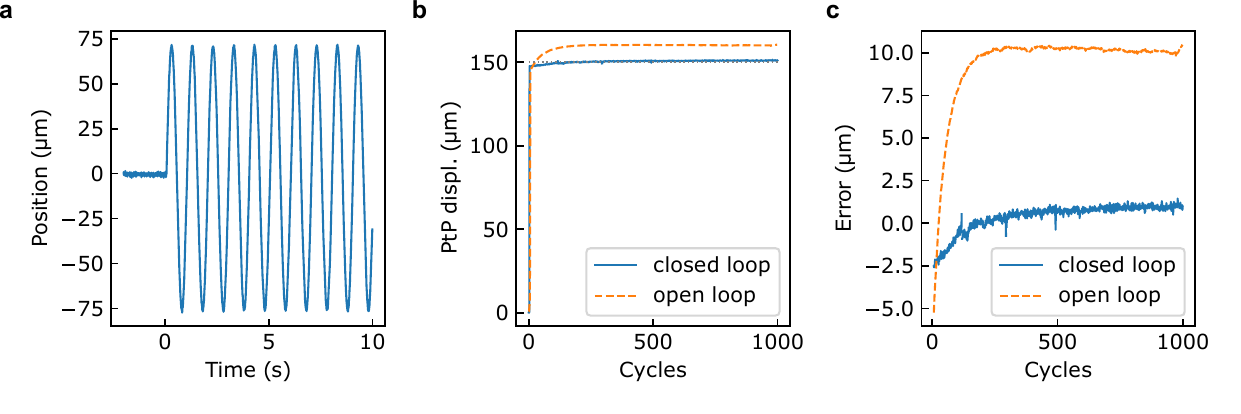}
\caption{\textbf{Temperature-induced displacement drift under open and closed-loop control.} \textbf{a} Displacement over time for the first 10 cycles. \textbf{b} Peak-to-peak displacement over 1000 cycles for open-loop and closed-loop operation. \textbf{c} Error (measured displacement - target displacement) over 1000 cycles for open-loop and closed-loop operation.}\label{fig:longterm}
\end{figure}

To evaluate the performance of the closed-loop control in compensating for mechanical drift, we performed a temperature stability test. The actuator was driven with a sinusoidal signal at \SI{1}{\hertz} with a target displacement of \SI{150}{\micro\meter} (\textpm \SI{75}{\micro\meter}) for \SI{1000}{cycles}/\SI{1000}{\second} which led to self-heating of the device. Figure \ref{fig:longterm}\,(a) traces the lens position monitored using the LDV over the first 10\,cycles. Figure\,\ref{fig:longterm}\,(b) plots the evolution of the peak-to-peak displacement as a function of time, while Fig.\,\ref{fig:longterm}\,(c) indicates the error between the measured displacement and the target displacement. For open-loop operation, the peak-to-peak displacement increased from \SI{145}{\micro\meter} (\textpm\SI{72.5}{\micro\meter}) to \SI{160}{\micro\meter} (\textpm\SI{80}{\micro\meter}), corresponding to \SI{10}{\percent} relative displacement increase overall. The time constant \(\tau\) of the error for open-loop operation of \SI{50}{\second} closely matches the temperature time constant of the coil pair (\(\tau_\textit{coil}\) = \SI{57}{\second}\,\textpm\,\SI{2}{\second} \cite{Lux2025}), confirming the temperature related drift of the mechanical response. It should also be noted that the initial displacement does not align with the target displacement, which further underscores the necessity for closed-loop control. In closed-loop operation, the drift in the measured peak-to-peak displacement over time is still present with a similar time constant, but is reduced by \SI{83}{\percent}.

The presence of the residual drift indicates that not all temperature-related mechanisms are fully corrected by the present sensing and control approach. Even with internally temperature-compensated Hall sensing, the field-to-position relationship can remain temperature dependent due to changes in the magnetic circuit and geometry. In particular, the remanence of the NdFeB magnet decreases with temperature (\SI{-0.115}{\percent\per\celsius}), altering the magnetic field at the sensor for a fixed displacement when the temperature is changed. Moreover, thermal expansion and temperature-dependent softening of surrounding components can slightly change the relative alignment between magnet and sensor. These effects effectively introduce a temperature dependence in \(\frac{d\vec{B}(T)}{dz_\textit{magnet}}\), which is not captured by a single static calibration and likely accounts for the remaining long-term drift observed under closed-loop operation.

\section{Discussion}\label{discussion}

This work demonstrates closed-loop control of a 3D nano-printed electromagnetic actuator using integrated magnetic position sensing. We show that reliable position readout can be achieved in the quasi-static regime by separating coil and magnet field contributions through calibration-based subtraction.

The precision of the demonstrated system is primarily limited by Hall sensor noise, as it follows the expected noise propagation through the calibrated field-to-position sensitivity. The precision can be improved by enabling the onboard digital filter of the Hall sensor, at the cost of a lower sampling rate. The observed accuracy exhibits a systematic error, which propagates into an offset between setpoint and actual displacement in subsequent experiments. Because the position estimate relies on subtracting a comparatively large coil-induced field and applying a linear field-to-position mapping, even small calibration or modeling errors translate directly into position bias. Consistent with this interpretation, the fit residuals from the linear calibration steps indicate small deviations from the linear model, suggesting that imperfect coil-field subtraction and weak nonlinearity in the field–position relationship contribute to the observed systematic error. Another effect reducing accuracy is the temperature-dependent remanence of the micro-magnet and the temperature dependence of the Hall sensor. In addition, time-varying external magnetic fields could also interfere with the Hall readout, resulting in apparent displacement errors. This suggests that shielding or periodic offset updates may be beneficial in magnetically noisy environments. A definitive attribution of the residual systematic error is not feasible within the present experimental scope, as this error reflects the combined influence of calibration residuals, model assumptions, fabrication tolerances, and temperature effects. Future implementations could address this limitation through temperature-dependent calibration or improved thermal management of the magnetic circuit.

While the closed-loop control system can compensate for viscoelastic creep after a step input, reduce ringing, and thus shorten the settling time, critical damping is not possible as the dynamic performance of the control system is limited by the bandwidth of the utilized Hall sensor, which has a sampling rate of \SI{1.4}{\kilo\hertz}. This limitation makes the controller sampling frequency only about four times higher than the natural frequency of the actuator (\SI{362}{\hertz}). The low ratio between the sampling frequency and the system dynamics introduces a phase delay due to zero-order hold and computational delays inherent to discrete-time feedback systems. As a result, increasing the gain near the natural response of the actuator would lead to excitation rather than suppression of the ringing. Active damping of the ringing would require maintaining gain near the frequency of the fundamental mode while maintaining sufficient phase margin. Consequently, active suppression of the fundamental mode to achieve a faster settling time would require a sensor with a higher sampling rate, ideally exceeding \SI{3.6}{\kilo\hertz}. 

Overall, despite residual systematic bias and temperature-dependent drift, the demonstrated closed-loop magnetic sensing enables displacement control with mean sub-micron precision over a 
\textpm\SI{75}{\micro\meter} range while compensating creep and hysteresis that otherwise hinder quasi-static operation. Additionally, the proposed control system reduces temperature-induced displacement errors, improving positional stability under changing thermal conditions and during continuous actuation. To the best of our knowledge, this work represents the first demonstration of closed-loop control in a monolithically 3D nano-printed electromagnetic actuator using integrated magnetic self-sensing for quasi-static positioning, providing a practical pathway toward robust use in miniaturized optical systems where repeatable, stable positioning is required.

\section{Materials and methods}\label{methods}

\subsection*{Manufacturing of the device}

\begin{figure}[htb]
\centering
\includegraphics[width=0.9\textwidth]{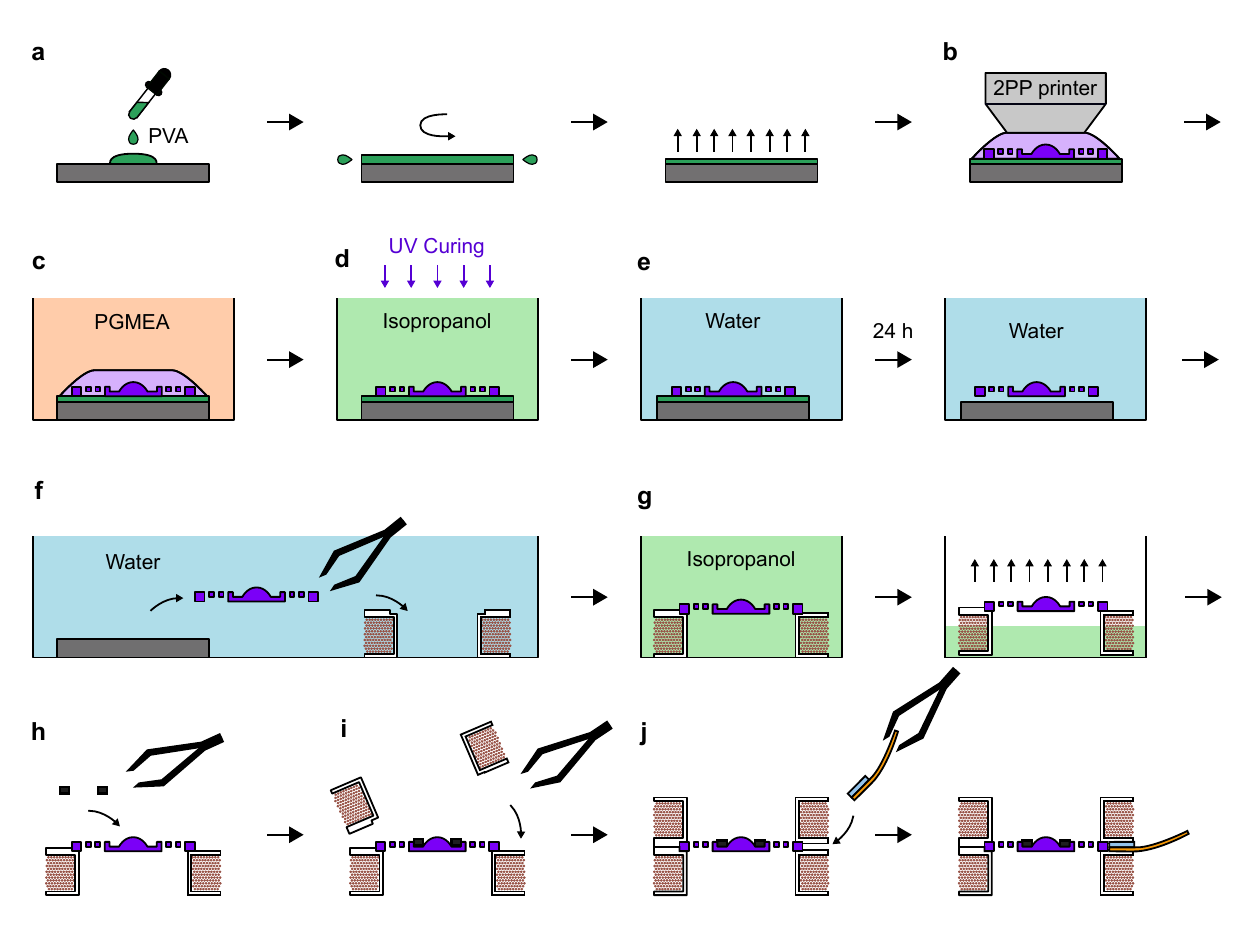}
\caption{\textbf{Fabrication of the actuator.} \textbf{a} A PVA layer of \SI{20}{\nano\meter} is spin-coated as a release layer. \textbf{b} The actuator is printed using two-photon polymerization and \textbf{c} developed in PGMEA. \textbf{d} While rinsed with Isopropanol, the actuator is illuminated with UV. \textbf{e} The actuator is then released using water to dissolve the sacrificial PVA layer, and \textbf{f} transferred onto the lower coil. \textbf{g} Water is exchanged by Isopropanol, which is evaporated. \textbf{h} The ring micro-magnet is placed on the shuttle and glued in place. \textbf{i} The second coil is placed before inserting the Hall sensor \textbf{j} and fixing the individual components with glue. Drawings not to scale.}\label{fig:manufacturing}
\end{figure}

The manufacturing process, which was adapted from our previous work \cite{Lux2025} is summarized in Fig.\,\ref{fig:manufacturing}. A \SI{20}{\nano\meter} polyvinyl alcohol (PVA; 87–89\,\% hydrolyzed, M\textsubscript{w} 13,000–23,000, Sigma-Aldrich 363081) release layer was spin-coated onto a silicon substrate from a 1 wt\% aqueous solution at \SI{1750}{RPM} using static dispense. The actuator was fabricated using a 3D nano-printer (Photonic Professional GT+, Nanoscribe GmbH, Germany) with a 10×/0.3 NA objective and IP-S photoresin in dip-in configuration. An overview of the printing parameters for the mechanical support, springs, and lens is provided in Tab\,\ref{tab:printing_parameters}. Mechanical components were printed using a large slicing distance to reduce printing duration. The lens was fabricated using a two-stage strategy: a coarse bulk print with large slicing distance, followed by a \SI{5}{\micro\meter} thick surface shell printed with reduced slicing distance to obtain a surface with low roughness. The overhanging springs were fabricated using a block-wise printing strategy adapted from \cite{Marschner2023}. Springs were subdivided into sequential parallelepipeds printed along the radial direction with a block length of \SI{20}{\micro\meter}, shear angle of \SI{15}{\degree}, and an overlap of \SI{4}{\micro\meter}. Toolpaths were generated using a custom General Writing Language (GWL) Python script. To prevent deformation during development, the springs were constrained by safety pins (height \SI{20}{\micro\meter}, width \SI{5}{\micro\meter}). After printing, the samples were developed in PGMEA for \SI{2}{\hour}, rinsed in IPA for \SI{30}{\minute}, and exposed at \SI{365}{\nano\meter} (LED Pen 2.0, Dr. Hönle AG) while submerged in IPA to increase cross-linking \cite{Schmid:19,Purtov2018}. Lift-off was achieved by immersing the substrate in water to dissolve the PVA layer. While submerged, the actuator was transferred onto the lower electromagnetic coil. Coil holders were fabricated from fused silica by selective laser-induced etching (LightFab GmbH, Germany). Coils were wound from enameled copper wire (TRU Components 1570225) using orthocyclic winding. Water was subsequently exchanged with IPA prior to drying. The safety pins were removed using multiphoton ablation \cite{Xiong2012} with the same laser system that was used for selective laser-induced etching. The system was used with a 20×/0.45 NA objective, an average power of \SI{100}{\milli\watt}, and a scan speed of \SI{80}{\micro\meter\per\second}. The micro-magnet, machined from pressed and sintered NdFeB VAC745HR (Audemars Microtec, Switzerland), was placed on the shuttle and fixed using UV-curable adhesive (Vitralit\textsuperscript{\textcopyright} UC 4731, Panacol-Elosol GmbH, Germany). The second coil was aligned using passive alignment structures and fixed with the same adhesive. The Hall sensor was mounted onto the flexible PCB using a reflow soldering process. The sensor was then inserted into its allocated groove and fixed in place using UV-curable adhesive.

\begin{table}[htb]
\caption{Printing parameters for fabricating the actuator in IP-S photoresin using a Nanoscribe Photonic Professional GT+ with a 10×/0.3 NA objective.}\label{tab:printing_parameters}%
\begin{tabular}{@{}lllll@{}}
\toprule
 Parameter & support  & springs & lens & lens shell\\
\midrule
slicing distance (\SI{}{\micro\meter})   & 1.5 & 2 & 1.5 & 0.2  
\\

hatching distance (\SI{}{\micro\meter}) &  0.5  & 1.0 & 0.5 & 0.5            \\
power (\SI{}{\milli\watt})            & 54.5 & 50 & 54.5  &54.5          \\
mode                & \multicolumn{4}{c}{solid}        \\
\bottomrule
\end{tabular}
\end{table}

\subsection*{Closed-loop control}\label{sec:closedloop}
The Hall sensor was read out over I2C with 
a Teensy\textsuperscript{\textcopyright} 4.0 USB development board (PJRC, OR). The closed-loop control was run on the development board. The current source is a custom bidirectional voltage-controlled current source/sink based on a 16-bit I2C digital-to-analog converter (AD5693RARMZ-RL7, Analog Devices, MA), two high-voltage, low-noise, zero-drift operational amplifiers (LTC2057HV, Analog Devices, MA), used for voltage subtraction and the feedback loop. The load is driven by a NPN/PNP high power double bipolar transistor (PHPT610030NPK, Nexperia, Netherlands). A high-precision chip resistor with a low temperature coefficient of resistance of \textpm\,\SI{0.2}{ppm\per\celsius} (Y402310R0000C9R, VPG Foil Resistors, PA) is used as a shunt resistor. 
The current source was calibrated over a range of \textpm\,\SI{75}{\milli\ampere}. The maximum error of the calibrated current source was \SI{0.86}{\micro\ampere}, corresponding to an error of \SI{0.0011}{\percent}.

\begin{figure}[htb]
\centering
\includegraphics[width=0.9\textwidth]{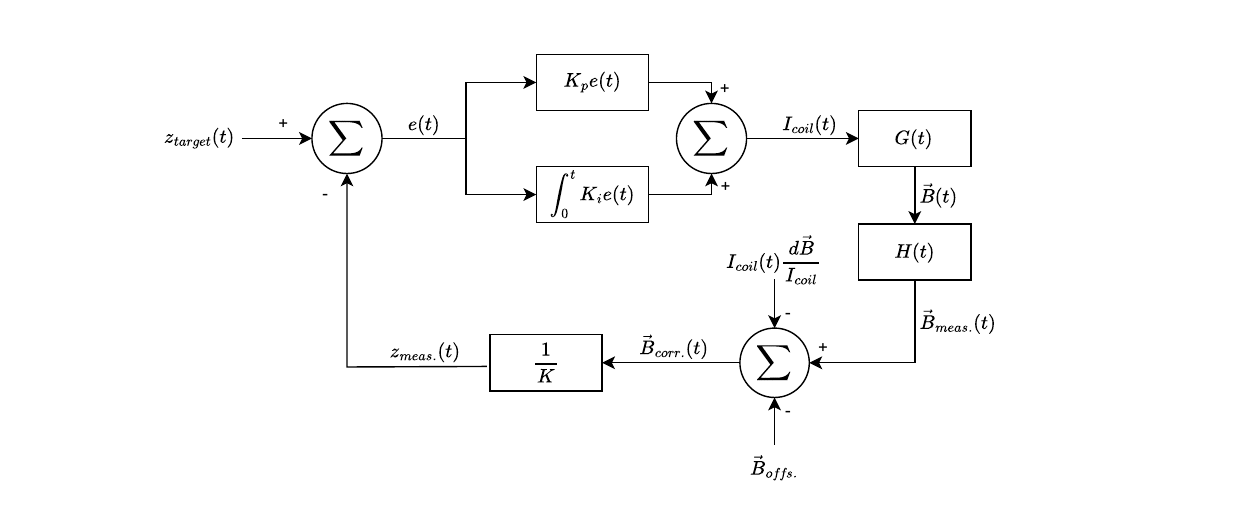}
\caption{\textbf{Closed-loop control.} A PI controller updates the coil current \(I_\textit{coil}\) from the displacement error \(e(t)=z_\textit{target}-z_\textit{meas.}(t)\). The Hall sensor measures the magnetic field \(\vec{B}\) created by the device \(G(t)\) and the coil contribution \(I_\textit{coil}\,\frac{d\vec{B}}{dI_\textit{coil}}\) and the Hall sensor offset \(\vec{B}_\textit{offset}\) are subtracted using calibration. The displacement is estimated as \(z_\textit{magnet}=\frac{B_y}{K}=B_y\left(\frac{dB_y}{dz_\textit{magnet}}\right)^{-1}\).}
\label{fig:closedloop}
\end{figure}

The block diagram of the closed-loop control scheme is illustrated in Fig.\,\ref{fig:closedloop}. A desired actuator displacement
\(z_\textit{target}\) is specified and compared with the measured displacement
\(z_\textit{meas.}(t)\), yielding the control error \(e(t)=z_\textit{target}-z_\textit{meas.}(t)\). This error signal is processed by a PI controller, which computes the updated coil current \(I_\textit{coil}(t)\). 
The applied coil current generates a magnetic field at the actuator, described by the system transfer function \(G(t)\). The resulting magnetic flux density $\vec{B}(t)$ is registered by the Hall sensor \(H(t)\). The measured field \(\vec{B}_\textit{meas.}(t)\) is corrected by subtracting the sensor offset $\vec{B}_\textit{offset}$ and the contribution of the coil to the magnetic field
\(I_\textit{coil}\) \(\frac{d\vec{B}}{d I_\textit{coil}}\) using calibration
coefficients. The actuator displacement is then estimated from the corrected magnetic field using the y-component according to \(z_\textit{meas.} = \frac{B_\textit{y, corr.}}{K}\) with \(K\)\ being \(\frac{d B_y}{d z_\textit{magnet}}.\) 

The process to calculate the estimated magnet displacement \(z_\textit{meas.}(t)\) can be written as

\begin{equation} \label{eq2}
z_\textit{meas.}(t) = \frac{B_y-B_\textit{offset,y}-I_\textit{coil}(t)\left(\frac{dB_\textit{y}}{dI_\textit{coil}}\right)}{\left(\frac{dB_\textit{y}}{dz_\textit{magnet}}\right)}.
\end{equation}

This estimation and control cycle is executed continuously, thereby regulating the
actuator displacement to the desired displacement.

\subsection*{LDV measurements}
A single-point LDV (VGo-200, Polytec GmbH, Germany) was used to measure the displacement data that were used as a reference measurement.

\subsection*{Simulation}
The magnetic field data used for comparing experimental data with simulated data were created using finite element analysis in the COMSOL Multiphysics \textsuperscript{\tiny\textregistered} AC/DC module.

\section*{Acknowledgments}

Funded by the Deutsche Forschungsgemeinschaft (DFG, German Research Foundation) – 555645878.

\subsection* {Code, Data, and Materials Availability} 

Data underlying the results presented in this paper are not publicly available but may be obtained from the authors upon reasonable request.

\section*{Conflict of interests}

The authors declare that there are no financial interests, commercial affiliations, or other potential conflicts of interest that could have influenced the objectivity of this research or the writing of this paper.

\section*{Author Contributions}

F.L. designed and fabricated the actuators and driving electronics. E.D. implemented and validated the closed-loop system on the development board. F.L. performed the experiments and analyzed the data. C.A. supervised this work and provided technical and theoretical support. F.L. drafted the manuscript. C.A. and E.D. revised the manuscript.

\bibliography{sn-bibliography}

\end{document}